\documentclass[twocolumn]{aastex61}

\submitjournal{ApJ}

\shorttitle{Jet Power of AGNs}
\shortauthors{Fan \& Wu}

\begin{document}

\title{Jet Power of Jetted Active Galactic Nuclei: Implication for Evolution and Unification}

\email{fanxl@sues.edu.cn}

\author[0000-0003-0988-9910]{Xu-Liang Fan}
\affil{School of Mathematics, Physics and Statistics, Shanghai University of Engineering Science, Shanghai 201620, China}
\affil{School of Physics, Huazhong University of Science and Technology, Wuhan 430074, China}

\author{Qingwen Wu}
\affiliation{School of Physics, Huazhong University of Science and Technology, Wuhan 430074, China}



\begin{abstract}
  We construct samples of jetted active galactic nuclei (AGNs) with low-frequency radio data from the recent released TGSS AD1 catalog at 150 MHz. With these samples, we compare the properties of jet power for blazars, radio-loud narrow-line Seyfert 1 galaxies (RL-NLS1s), young radio sources (YRSs) and radio galaxies. The jet-disk connection, and the unification of jetted AGNs are explored. On the Eddington ratio --- Eddington-scaled jet power plane, jetted AGNs can be generally divided into two populations. Low power radio galaxies, low excitation FR IIs (LEG/FR IIs), and most YRSs show larger jet power than accretion power, while FSRQs, RL-NLS1s, and high excitation FR IIs (HEG/FR IIs) are on the contrary. LEG/FR IIs share similar jet power properties with HEG/FR IIs, while their accretion properties are different with the latter. These facts suggest an evolutional sequence from HEG/FR IIs, LEG/FR IIs to FR Is, where the accretion and jet activities get dimmed gradually. LEG/FR IIs are the transitional objects that accretion processes have switched off, while jets are still active. After correcting the contribution from radio core of blazars, the unification between blazars and radio galaxies is confirmed with the jet power distributions. The unification involved RL-NLS1s is more complicated than the simple scenarios of black hole growth or orientation effect. In addition, our results manifest that low synchrotron peaked BL Lacs (LBLs) contain two distinct groups on the distribution of jet power, with one group similar with FSRQs and the other similar with intermediate synchrotron peaked BL Lacs. The LBLs with higher jet powers might be the aligned counterparts of LEG/FR IIs.
\end{abstract}




\section{Introduction} \label{sec:intro}
A small fraction of active galactic nuclei (AGNs) host relativistic jets (e.g.,~\citealt{2016ApJ...833...30C, 2017A&ARv..25....2P}). These jetted AGNs are usually found to be radio loud. When the jets move towards the earth, the radiation of AGNs is dominated by the non-thermal emission from aligned relativistic jets. These make up a subclass of AGNs, called blazars. When the viewing angles are larger, the Doppler boosting effect becomes insignificant, jetted AGNs are appeared as radio galaxies or steep-spectrum radio quasars (SSRQs). This scenario based on the orientation is usually called the unification model~\citep{1995PASP..107..803U}. The unification model considers that the two subclasses of blazars, BL Lacs and FSRQs, are unified with FR I radio galaxies (FR Is) and FR II radio galaxies (FR IIs), respectively.

The FR dichotomy of radio galaxies is mainly based on the morphologies of the large-scale radio structure~\citep{1974MNRAS.167P..31F}. FR Is are found to be edge-darkened, while FR IIs are edge-brightened. Following the initially morphology based classification, it was suggested that the luminosity and accretion mode are also distinct between FR Is and FR IIs~\citep{2004MNRAS.351..733M}, similar between BL Lacs and FSRQs~\citep{2009MNRAS.396L.105G, 2011MNRAS.414.2674G}. FR IIs and FSRQs are accepted to accrete at radiatively efficient state, while FR Is and BL Lacs are associated with radiatively inefficient accretion. There are also several suggestions that jetted AGNs with high and low accretion rates also correspond to high excitation radio galaxies (HEGs) and low excitation radio galaxies (LEG,~\citealt{2010A&A...509A...6B, 2012MNRAS.421.1569B}, but also see~\citealt{2016RAA....16..136H}), respectively. In observations, FR Is are usually found to be LEGs, while FR IIs contain both LEGs and HEGs~\citep{2010A&A...509A...6B, 2017A&A...598A..49C, 2017A&A...601A..81C}.

The spectral energy distributions (SEDs) of blazars are characterized by two bumps. Based on the peak frequency of synchrotron radiation, blazars can be divided into three subclasses, low synchrotron peaked blazars (LSPs), intermediate synchrotron peaked blazars (ISPs) and high synchrotron peaked blazars (HSPs;~\citealt{2010ApJ...716...30A, 2016ApJS..226...20F}). FSRQs are usually found to be LSPs, while BL Lacs with different SED classes, i.e., LSP BL Lacs (LBLs), ISP BL Lacs (IBLs), and HSP BL Lacs (HBLs) are all found. Recently,~\citet{2012MNRAS.420.2899G} explored the intrinsic properties of blazars based on the Monte-Carlo simulations, and proposed that LBLs were the results of selection effects and they were actually FSRQs with the emission lines diluted by the jet radiation~\citep{2011MNRAS.414.2674G, 2014MNRAS.445...81S}.

Another subclass of jetted AGNs which is thought to be important on the unification is radio-loud narrow-line Seyfert 1 galaxies/quasars (RL-NLS1s,~\citealt{2017FrASS...4....6F}). More and more evidences support that RL-NLS1s host powerful relativistic jets with smaller black hole mass, and will evolve to FSRQs or classical FR IIs~\citep{2009ApJ...707L.142A, 2015A&A...575A..13F, 2019ApJ...872..169P}. What is the parent population of flat-spectrum RL-NLS1s then becomes another problem. Some authors suggested that compact steep sources (CSSs, also know as young radio sources or YRSs) were another candidate type in addition to steep-spectrum RL-NLS1s~\citep{2016A&A...591A..98B, 2018A&A...614A..87B}.

Jet power is a fundamental parameter of jet physics, and less affected by the orientation effect. It is also found to be tightly connected with black hole mass and accretion process, which is known as the jet-disk connection (\citealt{2014MNRAS.445...81S, 2017ApJ...840...46I}, but also see~\citealt{2014MNRAS.440..269M}). Thus jet power is useful to clarify the unification and evolution on jetted AGNs. The estimation of jet power is usually based on the observations of radio lobes or hot spots~\citep{2010ApJ...720.1066C, 2013ApJ...767...12G}~\footnote{Actually this is for the estimation of kinetic jet power. In this paper we focus on the kinetic jet power where the radiative component is not included (see e.g.,~\citealt{2008MNRAS.385..283C, 2018ApJ...861...97F}). For simplicity, we use jet power denoting kinetic jet power throughout this paper.}. Limited by the observations, this method can only be applied to several tens nearby radio galaxies. Based on the direct constraints on jet power, there are several empirical relations between jet power and 151 MHz radio luminosity~\citep{1999MNRAS.309.1017W, 2010ApJ...720.1066C, 2013ApJ...767...12G}. Recent years, several new catalogs of all sky surveys at low radio frequency (around 150 MHz) have been released~\citep{2017MNRAS.464.1146H, 2017A&A...598A..78I, 2017A&A...598A.104S}. These catalogs provide good opportunities to explore the jet power properties of jetted AGNs.

For blazars, the extended emission may be dominant even at low-frequency radio observations. Thus the estimation of jet power using monochromatic radio luminosity for blazars should be with caution~\citep{2011ApJ...740...98M}. With the recently released TGSS (TIFR Giant metrewave radio telescope Sky Survey) AD1 (First alternative data release) catalog~\citep{2017A&A...598A..78I},~\citet{2018ApJ...869..133F} showed that the low-frequency radio emission of blazars was the mixture of core and extended components~\citep{2013ApJS..206....7M}, and the extended jet emission of blazars at 150 MHz could be recovered with a correction factor 2.1. This correction makes it possible to estimate jet power and explore the jet-disk connection for larger sample of blazars. In this paper, we will show that this correction makes good unification on the jet power distributions between blazars and radio galaxies.

In this paper, we cross-match several samples of currently known jetted AGNs with the TGSS AD1. Benefited from the large samples of TGSS samples, we can compare the jet power properties between different SED subclasses of blazars, as well as between jetted AGNs with and without $\gamma$-ray detections. Then we verify the unification model, and the jet-disk connection combined the main types of jetted AGNs. In section 2, we describe the samples of jetted AGNs used in this work. Section 3 shows the jet power properties of different samples. The jet-disk connection, especially the Eddington-scaled jet power and jet production efficiency, are discussed in Section 4. In section 5, we discuss some implications of our results. The main conclusions are summarized in section 6. Throughout this paper, we use a $\Lambda$CDM cosmology with $H_0 = 71$ km s$^{-1}$ Mpc$^{-1}$, $\Omega_{\rm M}$=0.27, $\Omega_{\Lambda}$=0.73~\citep{2009ApJS..180..330K}.

\section{Sample Property}
In this paper, we use the low-frequency radio catalog of the TGSS AD1 at 150 MHz to estimate jet power. The TGSS survey covers 90\% of all sky between -53$^\circ$ and +90$^\circ$ declination. The median sensitivity limit is about 3.5 mJy beam$^{-1}$. The astrometric accuracy is about 2\arcsec~in right ascension and declination~\citep{2017A&A...598A..78I}. The TGSS AD1 contains 0.62 million radio sources, which is one of the largest and deepest low frequency radio catalog currently~\citep{2017A&A...598A.104S}.

For blazars, we consider the multi-wavelength sample --- Roma-BZCAT~\citep{2009A&A...495..691M,2015Ap&SS.357...75M}, and the $\gamma$-ray sample --- the third catalog of AGNs detected by Fermi/LAT (3LAC) clean sample~\citep{2015ApJ...810...14A}. For RL-NLS1s, the $\gamma$-ray detected NLS1s~\citep{2019ApJ...872..169P} and the NLS1 sample selected from SDSS DR12~\citep{2017ApJS..229...39R} are applied in our analyses. The sample of YRSs is derived from a combined sample (most are CSS, 82/147) in Liao et al (2019, submitted, private communication). For radio galaxies, the 3CR sample with $z <$ 0.3~\citep{2010A&A...509A...6B, 2011A&A...525A..28B} and the recently builded FRICAT~\citep{2017A&A...598A..49C} and FRIICAT~\citep{2017A&A...601A..81C}, which have optical spectrographic observations to constrain the accretion properties, are used in our work.

\begin{figure}
\includegraphics[angle=0,scale=.28]{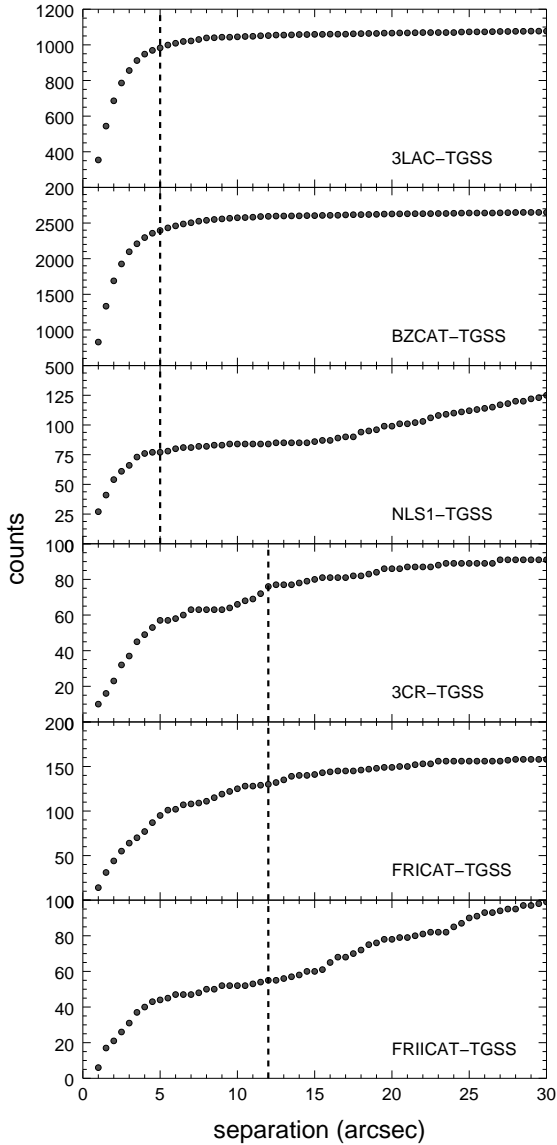}
\caption{The change of source counts with different matching radius. The panels from top to bottom are the results between 3LAC clean sample, BZCAT, DR12 NLS1s, 3CR, FRICAT, FRIICAT and TGSS AD1, respectively. The vertical dashed lines label the matching radius 5\arcsec~for the samples of blazars and NLS1s, and 12\arcsec~for the samples of radio galaxies. \label{sep}}
\end{figure}

Firstly, we compare the association results between various samples and TGSS AD1 for different matching radius. The changes of counts along with the separations are plotted in Figure~\ref{sep}. For blazars (3LAC clean sample and BZCAT), the counts are less variable when the radius is greater than 5\arcsec. The counts of the matching sample between DR12 NLS1s and TGSS AD1 also rise sharply when the separation is less than 5\arcsec. Interestingly, the counts get rise when the separation is larger than about 18\arcsec. This result may be related to the large-scale radio structures of RL-NLS1s (e.g.,~\citealt{2018ApJ...869..173R}), which is beyond the contents of this work. In order to avoid the spurious associations, we choose 5\arcsec~for the association of blazars and NLS1s. For the $\gamma$-ray NLS1s and YRSs, we find 15 of 16 objects in~\citet{2019ApJ...872..169P}, and 116 of 147 in Liao et al. (2019), respectively. The separations are all less than 10\arcsec. Thus we consider all these associations.

For radio galaxies, the conditions are a little completed. The counts change sharply when the radius is less than 5\arcsec, which reflect the position uncertainties between different observation instruments, just like the case for blazars. But as the radio galaxies is extended at radio band, the position could be different for different frequencies. For FR Is, the brightest component is generally coincident with the center of the host galaxy (SDSS coordinates). Thus the counts is less variable for FRICAT-TGSS sample when the radius is larger than about 15\arcsec. For FRIICAT-TGSS sample with separation greater than 15\arcsec, the counts are increasing due to two reasons. One is that the brightest components (hot spots) of FR IIs are not consistent with the optical cores. The other is that there are multiple TGSS objects for one FRIICAT object. We check these multiple associations, and find for the cases of two counterparts, the separations are generally equal, which may correspond to two radio lobes. Except for one case, one of the two separations is about 13\arcsec, smaller than the other association. When the matching radius is less than 12\arcsec, all the associations are unique. For 3CR-TGSS sample, the counts are less variable at 12\arcsec. In order to avoid the spurious associations, we choose 12\arcsec~for all the associations of radio galaxies~\footnote{We have tried several other matching radius and found that our results only change slightly.}.

Within the TGSS field, the clean sample of 3LAC has 1328 sources. We find 983 of them in TGSS AD1 within the separation of 5\arcsec. There are 3366 sources in BZCAT within the TGSS field. We find 2395 of them in TGSS AD1 within the separation of 5\arcsec. The detection rate is about 74.0\% and 71.2\% respectively, about two times larger than the previous works~\citep{2016A&A...588A.141G}.
Among the 3LAC-TGSS sample, there are 603 sources with redshift measurements, including 364 FSRQs, 197 BL Lacs and 42 AGNs of other types. Of the 364 FSRQs, there are 360 with the estimation of synchrotron peak frequency. The sample of BL Lacs contains 57 LBLs, 42 IBLs, 93 HBLs and 5 BL Lacs without peak frequency estimations. Among BZCAT-TGSS sample, there are 1987 sources with redshift measurements, including 1485 FSRQs, 222 BL Lacs, and 280 blazars of uncertain types.

77 NLS1s are associated within the separation of 5\arcsec. The detection rate is only 0.69\%. The jet power of NLS1s is generally larger than 10$^{43}$ erg s$^{-1}$, thus we consider they are all radio loud.~\citet{2017ApJS..229...39R} found 555 associations (detection rate 5.0\%) compared SDSS DR12 NLS1s with FIRST survey within a search radius of 2\arcsec. The deviation of the detection rates between these two radio bands can be due to the low sensitivity of low-frequency radio survey, or the relative flatter spectrum of RL-NLS1s between 150 MHz and 1.4 GHz. For YRSs, there are several objects with blazar-like features which have been labelled in Liao et al. (2019). These objects are excluded from our analyses. Finally there are 108 YRSs whose jet powers are estimated.

The 3CR sample contains 113 sources with $z < 0.3$, which have 76 common sources within 12\arcsec~of TGSS AD1 sources. We compare the luminosity between 3CR and TGSS AD1 for the 76 common sources. Most sources distribute along the equation line, only 5/76 objects show the dispersions larger than 0.3 dex. The mean value of the dispersion between 3CR and TGSS is 0.05 dex with the standard deviation 0.32 dex (when the five obvious outliers are excluded, the dispersion is only 0.02 dex with the standard deviation 0.08 dex). For a larger sample, we use the 178 MHz luminosity in 3CR and converted it into 150 MHz with a spectral index 0.7 ($S_{\nu} \propto \nu^{-\alpha}$). There are some suggestions that the LEG FR IIs is totally different with the HEG and broad line FR IIs~\citep{1995PASP..107..803U, 2010ApJ...723.1119L, 2013A&A...560A..81B}. Thus for the FR IIs of 3CR sample, the high excitation and broad line radio galaxies (hereafter HEG/FR IIs) are considered separately with low excitation radio galaxies (hereafter LEG/FR IIs). It contains 22 FR Is, 47 HEG/FR IIs and 19 LEG/FR IIs in 3CR sample. For the sources in FRICAT and FRIICAT, there are 130 and 55 objects within the 12\arcsec~around the position of TGSS AD1. Note that FR IIs in FRIICAT are mostly LEGs (49/55).

In order to explore the jet-disk connection, it needs the properties of bolometric luminosity of accretion process and the central black hole mass. We cross match the BZCAT-TGSS sample with SDSS DR7 quasar sample in~\citet{2011ApJS..194...45S}. This results in a blazar sample of 361 objects with black hole mass and bolometric luminosity estimations. Among them, 337 objects are classified as FSRQs in BZCAT. The bolometric luminosity and black hole mass listed in~\citet{2011ApJS..194...45S} are used. Among 108 YRSs, 69 of them have estimations of black hole mass and bolometric luminosity. For RL-NLS1s, 77 objects all have black hole mass and bolometric luminosity estimations. For the radio galaxies, the [O{\sc iii}] luminosity is used to estimate the bolometric luminosity with the correction factor $L_{\rm bol} = 3500~L_{\rm [O III]}$~\citep{2012MNRAS.421.1569B}. The black hole mass estimations are given for FRICAT and FRIICAT. The black hole mass of 3CR radio galaxies are estimated with host galaxy magnitude as~\citet{2010A&A...509A...6B}~\footnote{For blazars and NLS1s, the black hole mass and bolometric luminosity are estimated based on the continuum flux of spectrographic observations. The continuum flux can be overestimated due to the jet emission, especially for blazars. We compare the bolometric luminosity estimated from continuum and [O {\sc iii}] luminosity. Only 104 of 361 SDSS blazars have the [O {\sc iii}] luminosity, mostly due to the larger redshift. The bolometric luminosity estimated with [O {\sc iii}] luminosity is slightly larger than that with continuum luminosity, with the mean deviation 0.23 dex and the standard deviation 0.50 dex. Thus we conclude that the bolometric luminosity and black hole mass estimated from optical continuum are not overestimated by the beamed jet emission statistically.}. Finally, there are 127 objects in FRICAT and 55 objects in FRIICAT. For 3CR sample, there are 12 FR Is, 40 HEG/FR IIs and 18 LEG/FR IIs. The construction of each sample is also included in Table~\ref{cor}.

\section{The Properties of Jet Power}
Throughout this paper, the jet power is estimated with the P$_{\rm j}$ --- L$_{151}$ relation in~\citet{2013ApJ...767...12G}
\begin{equation}
P_{\rm kin} = 3 \times 10^{44} (\frac{L_{151}}{\rm 10^{25}~W~Hz^{-1}~sr^{-1}})^{0.67} ~~\rm erg~s^{-1}.
\label{pext}
\end{equation}
To correct the combined core and extended radio emission of blazars, $L_{151}/2.1$ are used in Equation~\ref{pext} to estimate their jet powers~\citep{2018ApJ...869..133F}.

\begin{table}
\begin{center}
  \caption{The Sample Properties and the Results of K-S Test.}
  \setlength{\tabcolsep}{2pt}
  \scriptsize
  \label{cor}
  \begin{tabular}{lllllllll}
  \hline
  Parameter & \multicolumn{3}{c}{Sample A} & \multicolumn{3}{c}{Sample B} & \multicolumn{2}{c}{K-S Test} \\
  & class & N & mean & class & N & mean & $D$ & $Prob$ \\
  \hline
  $P_{\rm j}$ & LBL & 57 & 44.65 & FSRQ & 360 & 45.18 & 0.36 & 4.3e-6 \\
  $P_{\rm j}$ & LBL(H) & 33 & 45.12 & FSRQ & 360 & 45.18 & 0.21 & 0.13\\
  $P_{\rm j}$ & LBL(L) & 24 & 43.99 & IBL & 42 & 44.08 & 0.20 & 0.55 \\
  $P_{\rm j}$ & LBL(L) & 24 & 43.99 & $\gamma$-NLS1 & 15 & 44.59 & 0.51 & 0.01 \\
  $P_{\rm j}$ & IBL & 42 & 44.08 & $\gamma$-NLS1 & 15 & 44.59 & 0.49 & 0.006 \\
  \hline
  $P_{\rm j}$ & $\gamma$-FSRQ & 364 & 45.18 & FSRQ & 1485 & 45.15 & 0.06 & 0.22 \\
  $P_{\rm j}$ & $\gamma$-BL Lac & 197 & 44.06 & BL Lac & 222 & 44.08 & 0.10 & 0.19 \\
  $P_{\rm j}$ & $\gamma$-NLS1 & 15 & 44.59 & NLS1 & 77 & 44.22 & 0.30 & 0.18 \\
  \hline
  $P_{\rm j}$ & HEG/FRII & 47 & 45.07 & LEG/FRII & 19 & 45.11 & 0.21 & 0.51 \\
  $P_{\rm j}$ & HEG/FRII & 47 & 45.07 & FSRQ & 1485 & 45.15 & 0.18 & 0.10 \\
  $P_{\rm j}$ & 3CR FRI & 22 & 44.16 & BL Lac & 222 & 44.08 & 0.21 & 0.30 \\
  $P_{\rm j}$ & FRICAT & 130 & 43.45 & BL Lac & 222 & 44.08 & 0.60 & 1.4e-26 \\
  $P_{\rm j}$ & FRIICAT & 55 & 43.88 & BL Lac & 222 & 44.08 & 0.18 & 0.12 \\
  $P_{\rm j}$ & NLS1 & 77 & 44.22 & FRIICAT & 55 & 43.88 & 0.43 & 8.5e-6\\
  $P_{\rm j}$ & NLS1 & 77 & 44.22 & YRS & 108 & 44.93 & 0.48 & 1.3e-9\\
  $P_{\rm j}$ & NLS1 & 77 & 44.22 & FSRQ & 1485 & 45.15 & 0.68 & 2.0e-30\\
  \hline
  $P_{\rm j}/L_{\rm Edd}$ & HEG/FRII & 40 & -1.93 & LEG/FRII & 18 & -2.03 & 0.19 & 0.72\\
  $P_{\rm j}/L_{\rm Edd}$ & HEG/FRII & 40 & -1.93 & FSRQ & 337 & -2.06 & 0.16 & 0.28\\
  $P_{\rm j}/L_{\rm Edd}$ & HEG/FRII & 40 & -1.93 & FRIICAT & 55 & -2.56 & 0.60 & 5.5e-8\\
  $P_{\rm j}/L_{\rm Edd}$ & LEG/FRII & 18 & -2.03 & FRIICAT & 55 & -2.56 & 0.62 & 2.9e-5\\
  $P_{\rm j}/L_{\rm Edd}$ & 3CR FRI & 12 & -3.24 & FRICAT & 127 & -3.21 & 0.15 & 0.96\\
  $P_{\rm j}/L_{\rm Edd}$ & NLS1 & 77 & -1.11 & YRS & 69 & -1.48 & 0.27 & 0.006 \\
  $P_{\rm j}/L_{\rm Edd}$ & NLS1 & 77 & -1.11 & FRIICAT & 55 & -2.56 & 0.81 & 1.6e-19 \\
  $P_{\rm j}/L_{\rm Edd}$ & NLS1 & 77 & -1.11 & FSRQ & 337 & -2.06 & 0.68 & 2.0e-26 \\
  \hline
  $P_{\rm j}/L_{\rm bol}$ & HEG/FRII & 40 & -0.37 & LEG/FRII & 18 & 0.62 & 0.82 & 2.7e-08\\
  $P_{\rm j}/L_{\rm bol}$ & HEG/FRII & 40 & -0.37 & FSRQ & 337 & -1.42 & 0.75 & 6.1e-19\\
  $P_{\rm j}/L_{\rm bol}$ & HEG/FRII & 40 & -0.37 & FRIICAT & 55 & 0.45 & 0.61 & 2.4e-8\\
  $P_{\rm j}/L_{\rm bol}$ & LEG/FRII & 18 & 0.62 & FRIICAT & 55 & 0.45 & 0.24 & 0.39\\
  $P_{\rm j}/L_{\rm bol}$ & LEG/FRII & 18 & 0.62 & 3CR FRI & 12 & 1.02 & 0.50 & 0.04\\
  $P_{\rm j}/L_{\rm bol}$ & 3CR FRI & 12 & 1.02 & FRICAT & 127 & 0.27 & 0.72 & 7.9e-6\\
  $P_{\rm j}/L_{\rm bol}$ & NLS1 & 77 & -0.68 & YRS & 69 & -0.02 & 0.52 & 1.7e-9 \\
  $P_{\rm j}/L_{\rm bol}$ & NLS1 & 77 & -0.68 & FRIICAT & 55 & 0.45 & 0.71 & 3.0e-15 \\
  $P_{\rm j}/L_{\rm bol}$ & NLS1 & 77 & -0.68 & FSRQ & 337 & -1.42 & 0.54 & 5.1e-17 \\
  \hline
  $M_{\rm BH}$ & FRICAT & 127 & 8.54 & 3CR FRI & 12 & 9.09 & 0.74 & 4.6e-6\\
  $M_{\rm BH}$ & FRIICAT & 55 & 8.32 & HEG/FRII & 40 & 8.86 & 0.63 & 7.1e-9\\
  $M_{\rm BH}$ & FRIICAT & 55 & 8.32 & LEG/FRII & 18 & 9.02 & 0.80 & 1.4e-8\\
  $M_{\rm BH}$ & FRIICAT & 55 & 8.32 & NLS1 & 77 & 7.21 & 0.89 & 1.6e-23\\
  $M_{\rm BH}$ & NLS1 & 77 & 7.21 & YRS & 69 & 8.35 & 0.76 & 2.3e-19\\
  $M_{\rm BH}$ & NLS1 & 77 & 7.21 & FSRQ & 337 & 9.00 & 0.97 & 0.00\\
  \hline
  $L_{\rm bol}$ & NLS1 & 77 & 44.90 & FRIICAT & 55 & 43.43 & 0.81 & 9.2e-20 \\
  $L_{\rm bol}$ & NLS1 & 77 & 44.90 & YRS & 69 & 45.00 & 0.25 & 0.01 \\
  $L_{\rm bol}$ & NLS1 & 77 & 44.90 & FSRQ & 337 & 46.47 & 0.88 & 2.8e-44 \\
  $L_{\rm bol}/L_{\rm Edd}$ & NLS1 & 77 & -0.42 & FRIICAT & 55 & -3.01 & 0.97 & 6.9e-28\\
  $L_{\rm bol}/L_{\rm Edd}$ & NLS1 & 77 & -0.42 & YRS & 69 & -1.46 & 0.71 & 2.6e-17\\
  $L_{\rm bol}/L_{\rm Edd}$ & NLS1 & 77 & -0.42 & FSRQ & 337 & -0.64 & 0.25 & 7.4e-4\\
  \hline
  \end{tabular}
\end{center}
\end{table}

\subsection{Jet Power of $\gamma$-ray Sources}
\begin{figure}
\includegraphics[angle=0,scale=.4]{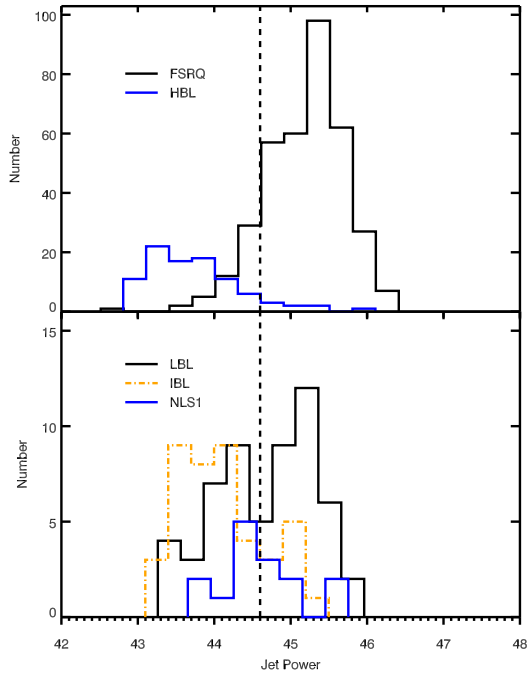}
\caption{The distributions of jet power for Fermi detected blazars and NLS1s. The vertical dashed lines label the jet power 10$^{44.6}$ erg s$^{-1}$ which divides the LBLs into two groups (see the text for details). The representation of each line is labelled in the figures. \label{3lac}}
\end{figure}

The jet power distributions of $\gamma$-ray blazars and NLS1s are plotted in Figure~\ref{3lac}. The jet power of LBLs shows a very broad distribution, which is somewhat bimodal. The K-S test between the distributions of FSRQs and LBLs shows clear distinct distributions, with $D = 0.36$ and $Prob = 4.3\times10^{-6}$ (also see Table~\ref{cor})~\footnote{Throughout this paper, we consider that the two populations are drawn from the same distribution when the $Prob$ is greater than 0.05, which do not reject the null hypothesis.}. The bimodal distribution of LBLs suggests that they may contain two populations, one is actually FSRQs as suggested by~\citet{2012MNRAS.420.2899G}. The other is transitional type BL Lacs which show weak emission lines and intermediate jet power.

We generally divide LBLs into two groups with the limit 10$^{44.6}$ erg s$^{-1}$ (Figure~\ref{3lac}). 33 of 57 LBLs with the jet power larger than 10$^{44.6}$ erg s$^{-1}$ show similar jet power distribution with FSRQs. The other LBLs show similar distribution with IBLs (Table~\ref{cor}). The mean values of the jet power of high and low power groups are also similar with FSRQs and IBLs, respectively (Table~\ref{cor}).

\begin{figure}
\includegraphics[angle=0,scale=.4]{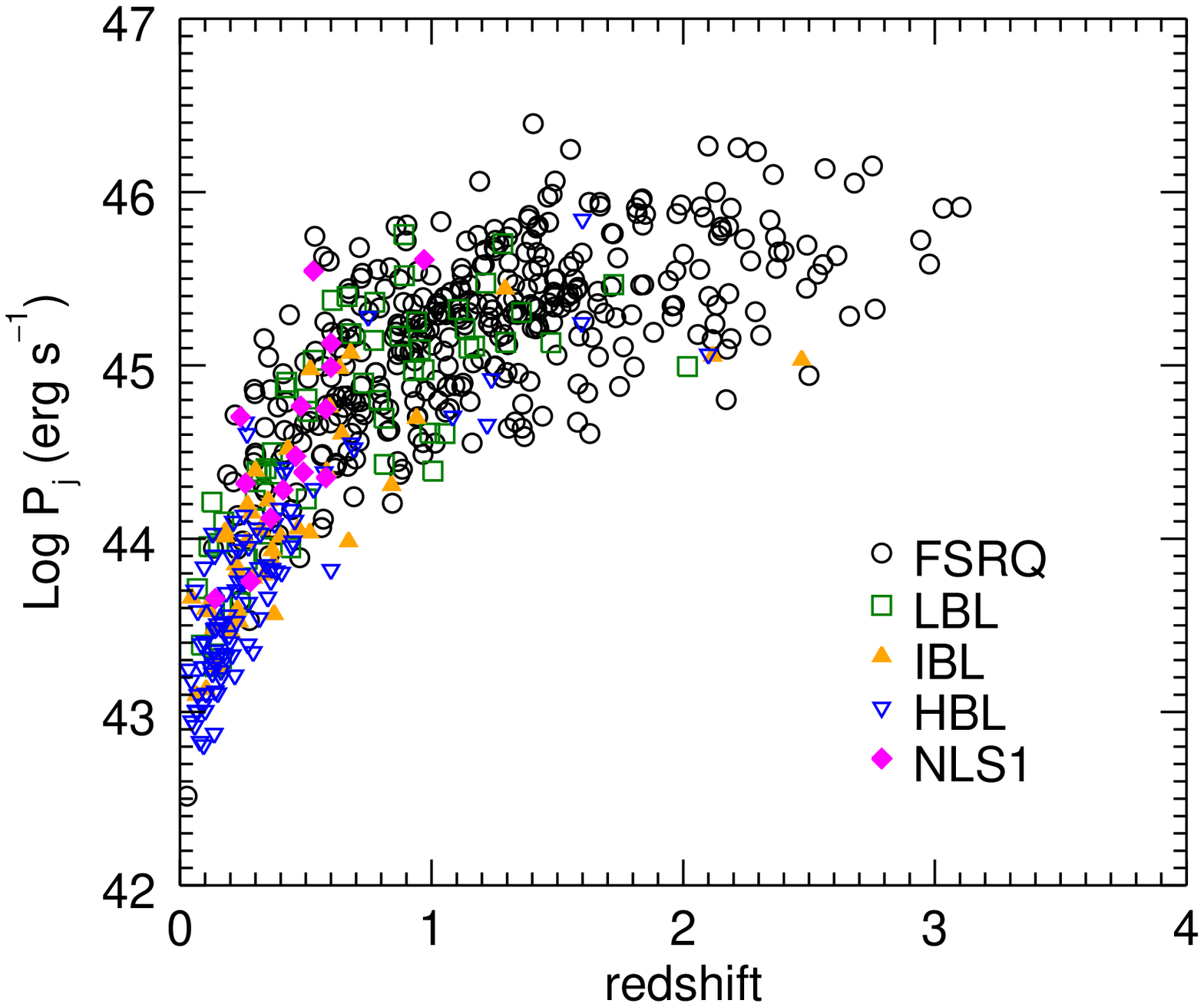}
\includegraphics[angle=0,scale=.4]{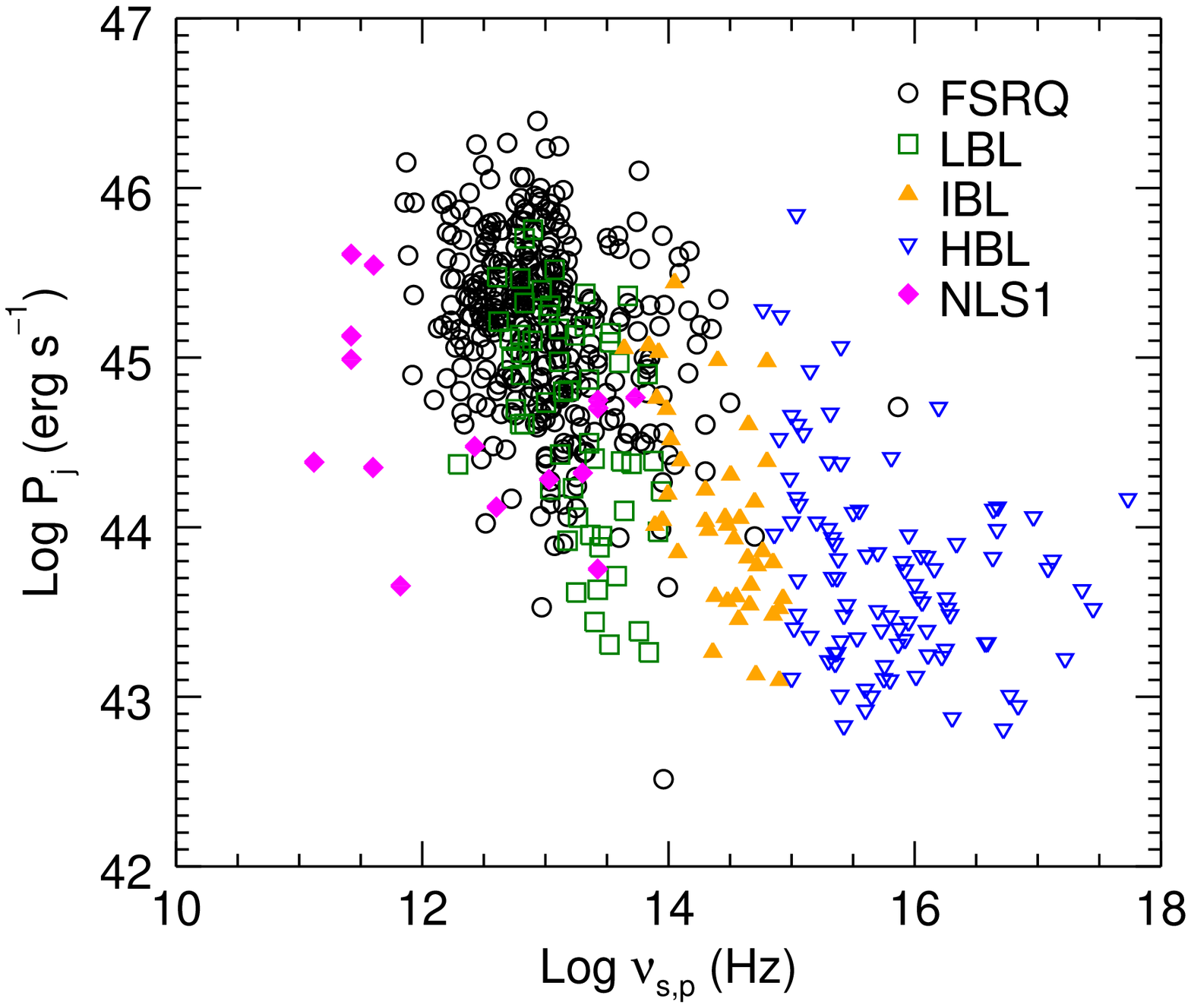}
\caption{Top panel: The jet power distribution for Fermi detected sources along with the redshift. Bottom panel: The jet power distribution along with the peak frequency.  The representation of each symbol is labelled in the figures. \label{pj_z}}
\end{figure}

The top panel of Figure~\ref{pj_z} shows the distribution of jet power along with the redshift. LBLs also show a wide range of redshifts, which is overlapped with both FSRQs and BL Lacs. The higher power LBLs locate at high redshifts similar with FSRQs, while the lower power ones locate at low redshifts, which are similar with BL Lacs.

The mean jet power of $\gamma$-ray NLS1s is smaller than FSRQs and high power LBLs, but larger than HBLs and IBLs (Table~\ref{cor}). The K-S tests show that their distribution of jet power is different with other classes of blazars.~\citet{2019ApJ...872..169P} estimated jet power of $\gamma$-ray NLS1s based on the results of SED fit, and found similar results.

\subsubsection{Blazar Sequence}
The well established blazar sequence is represented by a significant correlation between (bolometric) luminosity and peak frequency of blazars (\citealt{1998MNRAS.299..433F}, but also see~\citealt{2011ApJ...740...98M, 2012MNRAS.420.2899G}). Blazar sequence is usually explained as the increasing cooling from the brightening external photon field, which results in the lower peak and higher luminosity~\citep{1998MNRAS.301..451G}. Recent discoveries of $\gamma$-ray NLS1s and high-frequency high-luminosity blazars challenge this scenario~\citep{2012MNRAS.422L..48P, 2017FrASS...4....6F}. Here we examine the correlation between peak frequency and jet power in the bottom panel of Figure~\ref{pj_z}. These two parameters show an obvious negative trend, with the correlation coefficient of Spearman correlation test $\rho = -0.62$, and the chance probability of no correlation $P = 1.7\times10^{-64}$ for blazars. When NLS1s are included, this trend gets slightly better with $\rho = -0.64$ and $P = 6.5\times10^{-67}$. The peak frequency is related to radiation processes in the dissipation region at pc-scale, while jet power is associated with the physics of large-scale extended jet structures. Thus the correlation between these two parameters should not be direct, but indicates that the peak frequency is dependent on some fundamental properties of jet physics, such as the black hole mass~\citep{2016RAA....16..173F}, or the evolution of central engines~\citep{2017FrASS...4....6F}.

\subsection{Jet Power of $\gamma$-ray and non-$\gamma$-ray Sources}
\begin{figure}
\includegraphics[angle=0,scale=.4]{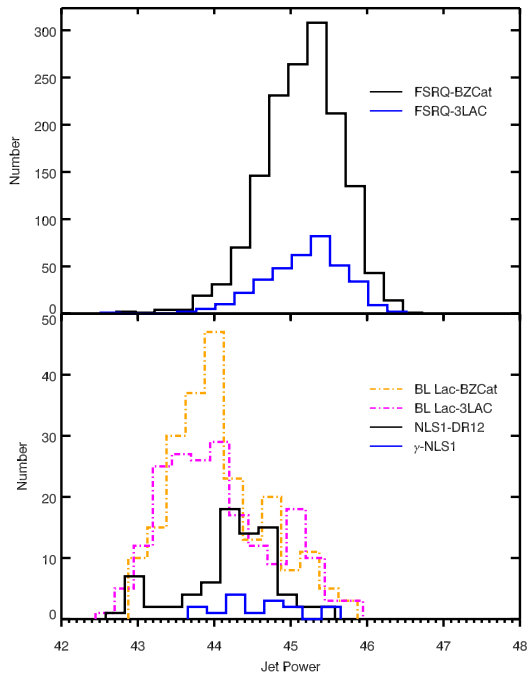}
\caption{The distributions of jet power for $\gamma$-ray and non-$\gamma$-ray sources.  The representation of each line is labelled in the figures.  \label{bzcat}}
\end{figure}

A fraction of blazars, as well as most RL-NLS1s are not detected by Fermi/LAT. Why these sources are $\gamma$-ray weak? Most researches found that the $\gamma$-ray sources had relatively higher Doppler factors~\citep{2009A&A...507L..33P, 2009ApJ...696L..22L, 2015ApJ...810L...9L, 2010A&A...512A..24S, 2014A&A...562A..64W}. Here we compare the jet powers of $\gamma$-ray blazars and NLS1s with those of the corresponding $\gamma$-ray quiet samples (Figure~\ref{bzcat}). $\gamma$-ray and non-$\gamma$-ray sources show no difference on the distributions of jet power (i.e., extended jet emission, Table 1). These results support that $\gamma$-ray and non-$\gamma$-ray sources share similar intrinsic properties on jet physics, and the difference between $\gamma$-ray and non-$\gamma$-ray sources is mainly caused by different viewing angles.

\subsection{Jet Power of Jetted AGNs}
\begin{figure}
\begin{center}
\includegraphics[angle=0,scale=.4]{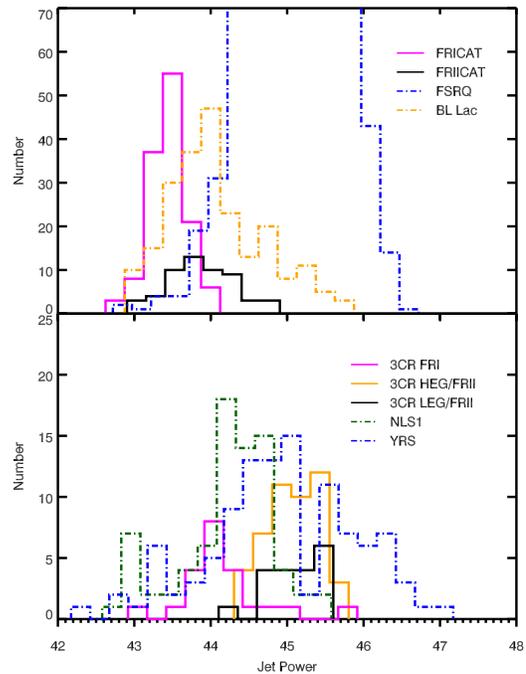}
\caption{The jet power distributions of jetted AGNs. The representation of each line is labelled in the figures. \label{rg}}
\end{center}
\end{figure}

Radio galaxies, as the unified version of blazars, should have similar distribution of jet power with blazars~\citep{1995PASP..107..803U}. We examine the jet power distribution of the BZCAT blazar sample and three samples of radio galaxies (Figure~\ref{rg}). To stress the radio galaxies with small sample sizes, we confine the peak to 80 in the upper panel of Figure~\ref{rg}, where the lacking distribution of FSRQs is same with that of Figure~\ref{bzcat}. BL Lacs and FR Is in 3CR have similar jet power, while FSRQs have similar jet power with HEG/FR IIs and LEG/FR IIs in 3CR (Table~\ref{cor}). The K-S test shows that BL Lacs and 3CR FR Is share similar distributions. In addition, FSRQs and HEG/FR IIs have similar jet power distributions, while HEG/FR IIs and LEG/FR IIs also belong to same distribution on jet power (Table~\ref{cor}).

The radio galaxies in FRICAT and FRIICAT have much lower jet power than 3CR radio galaxies (Table~\ref{cor}). Interestingly, the K-S test shows that BL Lacs have similar jet power distributions with FR IIs in FRIICAT, while BL Lacs and FRICAT objects are derived from distinct samples (Table~\ref{cor}).

The jet power of RL-NLS1 is generally less than FSRQs and classic FR IIs, and larger than that of BL Lacs and radio galaxies in FRICAT and FRIICAT. YRSs show a very broad distribution of jet power, from 10$^{42.17}$ erg s$^{-1}$ to 10$^{46.98}$ erg s$^{-1}$. Fractional YRSs show jet power similar with high-power blazars, while some of them are similar with NLS1s. The broad jet power distribution indicates that there contains evolution among different subclasses of YRSs, e.g., CSSs and GHz peaked sources. The classification of YRSs needs more detailed investigations.

\section{The jet-disk connection}
\subsection{The Connection between Jet Power and Accretion Power}
\begin{figure}
\includegraphics[angle=0,scale=.4]{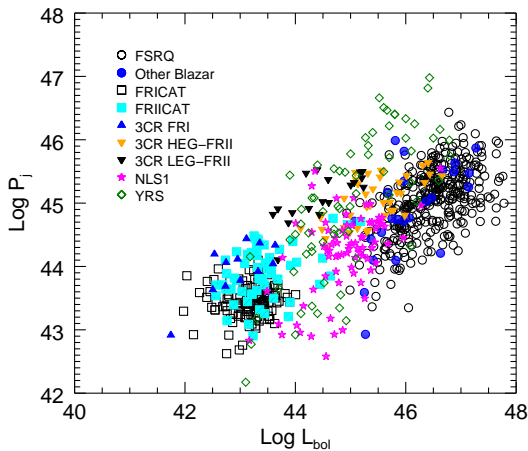}
\caption{The connection between accretion power and jet power. The representation of each symbol is labelled in the figure.  \label{lbol_pj}}
\end{figure}
Figure~\ref{lbol_pj} shows the scatters of accretion power and jet power. The partial Kendall's $\tau$ correlation test~\citep{1996MNRAS.278..919A} shows that the hypothesis of no correlation between these two parameters is rejected at a significance 0.05 when the influence from redshift is excluded, with $\tau = 0.28$ and $\sigma = 0.02$. FSRQs and HEG/FR IIs lie on the right-top corner on the accretion power --- jet power diagram. FR Is in 3CR and FRICAT as well as the FR IIs in FRIICAT stay on the left-bottom region with smaller jet power and accretion power. RL-NLS1s generally concentrate on the middle region, while YRSs span wide distributions on both accretion power and jet power. The results for blazars and RL-NLS1s are consistent with other works where the jet power was constrained by various methods~\citep{2014MNRAS.445...81S, 2017FrASS...4....6F, 2019ApJ...872..169P}

\subsection{The Distribution of Eddington-scaled Jet Power and Jet Production Efficiency}
In this section, we estimate Eddington-scaled jet power ($P_{\rm j}/L_{\rm Edd}$) and jet production efficiency ($P_{\rm j}/L_{\rm bol}$~\footnote{Note that the production efficiency is also defined as $\eta = P_{\rm j}/\dot{M} c^{2}$ (e.g.,~\citealt{2013ApJ...764L..24S}, where $\dot{M}$ is the accretion rate), which equals to $P_{\rm j}/L_{\rm bol}$ times the radiative efficiency of accretion process. These two definitions are equivalent if the radiative efficiency is considered as a constant.}) for various samples of jetted AGNs.
\begin{figure}
\includegraphics[angle=0,scale=.4]{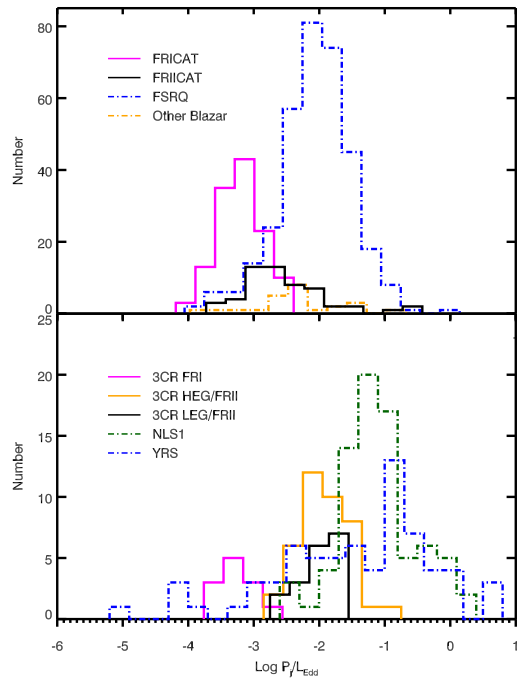}
\caption{The distributions of Eddington-scaled jet power  for jetted AGNs.  The representation of each line is labelled in the figures.  \label{pj_ledd_dis}}
\end{figure}
Figure~\ref{pj_ledd_dis} shows the distributions of Eddington-scaled jet power. The mean Eddington-scaled jet power of FSRQs, HEG/FR IIs and LEG/FR IIs are similar (Table~\ref{cor}), although the range of FSRQs is very broad and extends to $10^{-4}$. The K-S tests show that these three populations share similar distributions (Table~\ref{cor}). $P_{\rm j}/L_{\rm Edd}$ of FR Is in both 3CR and FRICAT are similar. The K-S test shows that these two populations share same distribution (Table~\ref{cor}). FR IIs in FRIICAT have intermediate Eddington-scaled jet power. The K-S tests show that their distribution is different from FR IIs in 3CR, including both HEGs and LEGs (Table~\ref{cor}). RL-NLS1s and YRSs own the highest Eddington-scaled jet power. But their distributions are completely distinct (Table~\ref{cor}). All RL-NLS1s have Eddington-scaled jet power larger than $10^{-3}$. The Eddington-scaled jet powers of YRSs span a wide range, while those of several YRSs are as low as $10^{-5}$.

\begin{figure}
\includegraphics[angle=0,scale=.4]{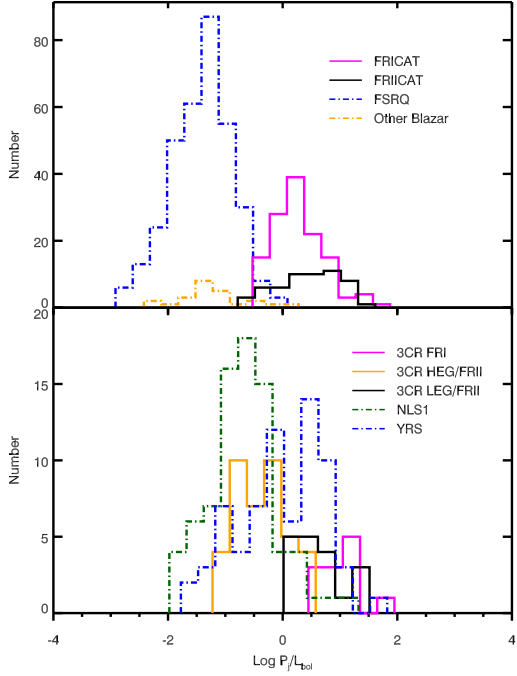}
\caption{The distributions of jet production efficiency for jetted AGNs.  The representation of each line is labelled in the figures. \label{pj_lbol_dis}}
\end{figure}
Figure~\ref{pj_lbol_dis} shows the distributions of jet production efficiency. FR Is in 3CR have the largest jet production efficiency. FR Is in FRICAT, FR IIs in FRIICAT, and LEG/FR IIs in 3CR show relatively high $P_{\rm j}/L_{\rm bol}$ with $P_{\rm j} > L_{\rm bol}$. The HEG/FR IIs, FSRQs, and RL-NLS1s have the lowest jet production efficiency. Again, YRSs show a wide distribution of $P_{\rm j}/L_{\rm bol}$ from 0.01 to 10, with the mean value close to 1. We also apply K-S tests for the distributions of jet production efficiency of various samples, only the LEG/FR IIs in 3CR show similar distribution with FR IIs in FRIICAT (Table~\ref{cor}).

\subsection{The Connection with Central Black Hole Mass}
\citet{1996AJ....112....9L} plotted the scatter between radio luminosity and absolute magnitude of host galaxy for radio galaxies, and found a clear division between FR Is and FR IIs on this diagram. This division is suggested as a switch at a fixed ratio between the jet power and Eddington luminosity (\citealt{2001A&A...379L...1G, 2008ApJ...687..156W}, also see the review of~\citealt{2016A&ARv..24...10T}).
\begin{figure}
\includegraphics[angle=0,scale=.4]{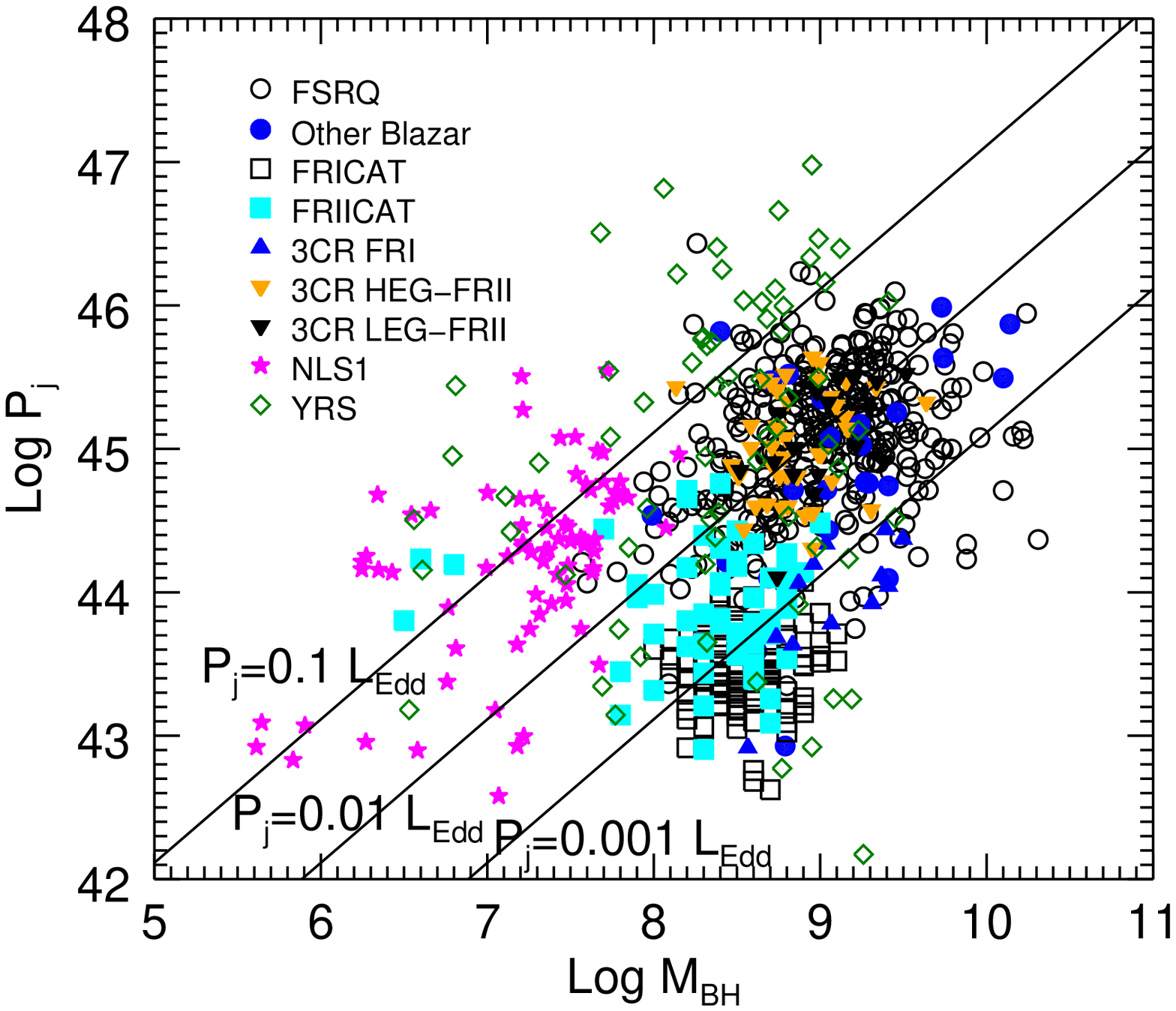}
\includegraphics[angle=0,scale=.4]{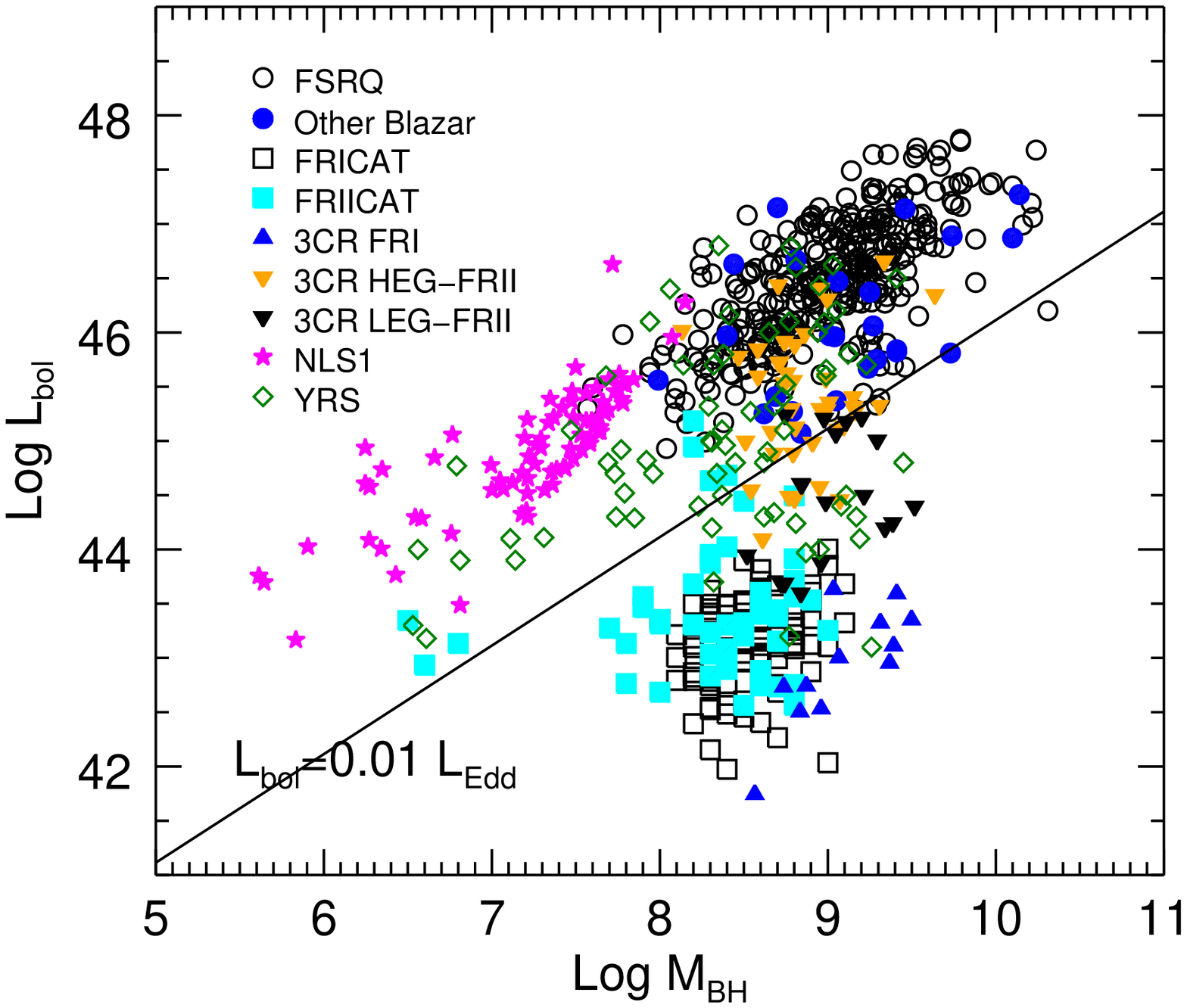}
\caption{Top panel: The jet power against the black hole mass for jetted AGNs. From high to low, the solid lines represent the jet power corresponding to $0.1~L_{\rm Edd}$, $0.01~L_{\rm Edd}$, and $0.001~L_{\rm Edd}$, respectively. Bottom panel: The bolometric luminosity of accretion process against the black hole mass for jetted AGNs. The solid line shows the bolometric luminosity corresponding to $0.01~L_{\rm Edd}$.  The representation of each symbol is labelled in the figures.  \label{pj_mbh}}
\end{figure}

In the top panel of Figure~\ref{pj_mbh}, we plot the jet power versus central black hole mass for the samples in our work. The ratio between jet power and Eddington luminosity 0.1, 0.01, and 0.001 are overplotted on the $P_{\rm j}$ --- $M_{\rm BH}$ plane. These lines seem unable to recover the division between FR Is and FR IIs/FSRQs. Actually, there seems no clear boundary or division line among different subclasses of jetted AGNs. This means the jet power is not only depend on the black hole mass, but also on some other physical features, such as the black hole spin, evolution, or the jet production efficiency, which is shown in Figure~\ref{pj_lbol_dis}.

There are also other suggestions that the FR Is and FR IIs/FSRQs can be divided on the $L_{\rm bol}$ --- $M_{\rm BH}$ plane~\citep{2010A&A...509A...6B, 2016A&ARv..24...10T}. The bottom panel of Figuire~\ref{pj_mbh} shows the bolometric luminosity versus central black hole mass for our samples. On the $L_{\rm bol}$ --- $M_{\rm BH}$ plane, the division is much clearer, with the division line around $L_{\rm bol}$ = 0.01 $L_{\rm Edd}$. This result confirms the division on the accretion mode for jetted AGNs. RL-NLS1s, FSRQs, most YRSs and HEG/FR IIs in 3CR are stay above the line, while FR Is and most LEG/FR IIs are distributing below the line. There are some FR IIs crossing the division line on the $L_{\rm bol}$ --- $M_{\rm BH}$ plane, including HEG/FR IIs and LEG/FR IIs in 3CR, as well as several FR IIs in FRIICAT.

As noted by~\citet{2017FrASS...4....6F}, jetted AGNs with low black hole mass and radiatively inefficient accretion are not observed on the left-bottom region of the $L_{\rm bol}$ --- $M_{\rm BH}$ plane. This can be a selection bias due to their low luminosity and low jet power.

Three FR IIs with extreme small black hole mass, which noted by~\citet{2017A&A...601A..81C} locate at the similar region with RL-NLS1s and YRSs on the $L_{\rm bol}$ --- $M_{\rm BH}$ and $P_{\rm j}$ --- $M_{\rm BH}$ plane. These three FR IIs have much smaller black hole mass, but relative higher bolometric luminosity and jet power.

\subsection{The Connection with Eddington Ratio}
\begin{figure}
\includegraphics[angle=0,scale=.4]{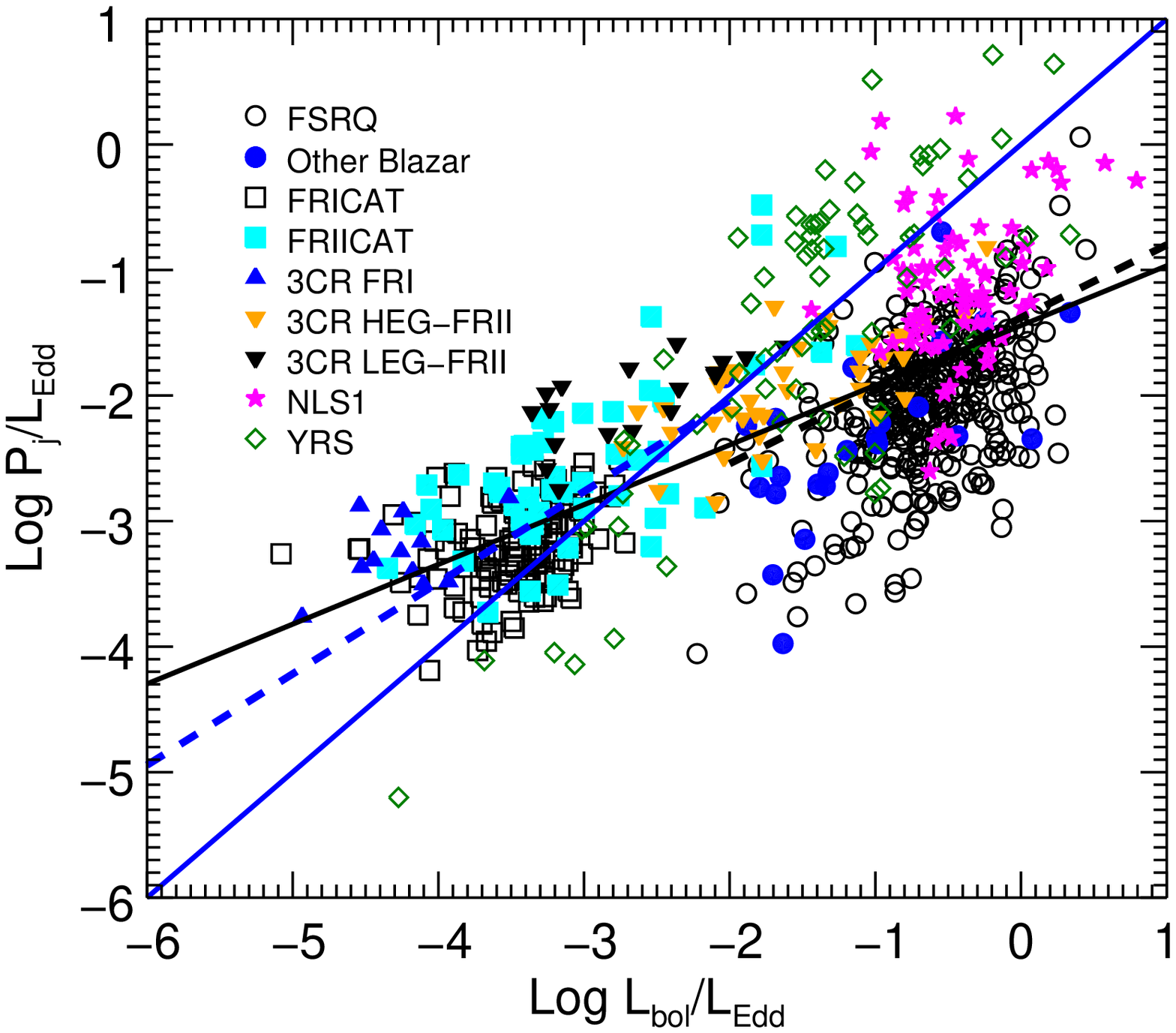}
\includegraphics[angle=0,scale=.4]{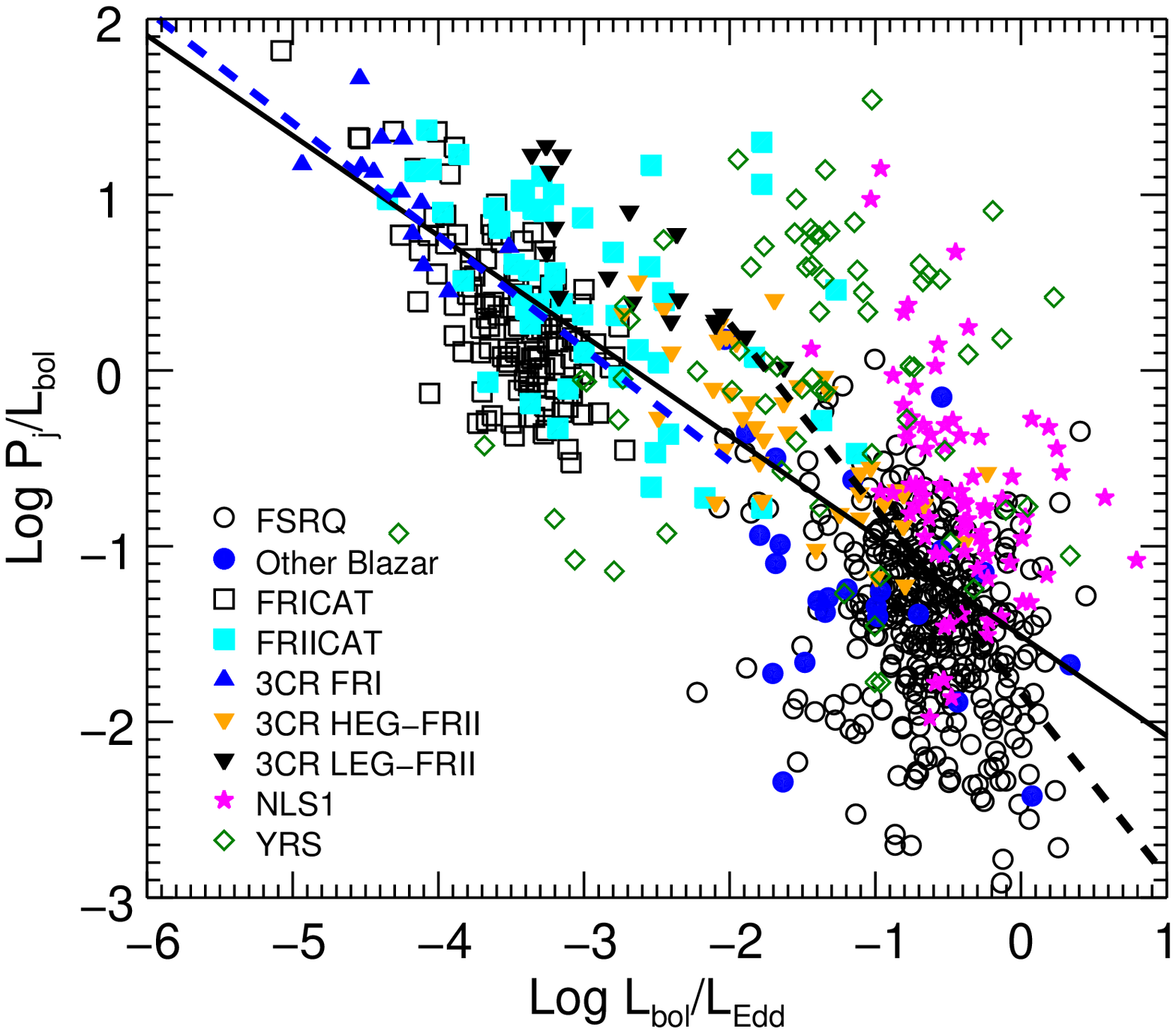}
\caption{Top panel: The Eddington ratio versus Eddington-scaled jet power of jetted AGNs. Bottom panel: The Eddington ratio versus jet production efficiency of jetted AGNs. The black solid line is the linear fit for the whole samples. The black and blue dashed lines are the linear fits for the sources with Eddington ratio larger and less than 0.01, respectively. The blue solid line in the top panel shows the equation line. The representation of each symbol is labelled in the figures.  \label{eddr_pj}}
\end{figure}

The top panel of Figure~\ref{eddr_pj} shows the correlation between Eddington-scaled jet power and Eddington ratio, which is also seen as a proxy of the jet-disk connection~\citep{2014MNRAS.445...81S}. The equation line (blue solid line) divide the sources into two general groups. When the Eddington ratio is relatively high, the accretion power is larger than jet power except for YRSs and fractional RL-NLS1s. When the Eddington ratio is low, generally less than 0.01, the jet power of most sources is larger than accretion power. The LEG/FR IIs and HEG/FR IIs in 3CR can generally be divided by the equation line on the $L_{\rm bol}/L_{\rm Edd}$ --- $P_{\rm j}/L_{\rm Edd}$ diagram. The $P_{\rm j}/L_{\rm Edd}$ of all the LEG/FR IIs is larger than $L_{\rm bol}/L_{\rm Edd}$, while most HEG/FR IIs, FSRQs and RL-NLS1s show smaller $P_{\rm j}/L_{\rm Edd}$ compared with $L_{\rm bol}/L_{\rm Edd}$. FR Is in 3CR, and most radio galaxies in FRICAT and FRIICAT with lower Eddington ratio have the Eddington-scaled jet power larger than accretion power. YRSs are the obvious outliers on the $L_{\rm bol}/L_{\rm Edd}$ --- $P_{\rm j}/L_{\rm Edd}$ diagram. Most YRSs have relatively high Eddington ratio, but their Eddington-scaled jet power is larger than Eddington ratio. We fit the linear relation for the sources with Eddington ratio larger and less 0.01. The results give that
\begin{eqnarray}
&&\log P_{\rm j}/L_{\rm Edd} = 0.73\pm0.08~\log L_{\rm bol}/L_{\rm Edd} - 0.59\pm0.25 \nonumber \\
&&(\sigma = 0.30~~{\rm for}~L_{\rm bol}/L_{\rm Edd} < 0.01), \\
&&{\rm and} \nonumber \\
&&\log P_{\rm j}/L_{\rm Edd} = 0.58\pm0.10~\log L_{\rm bol}/L_{\rm Edd} - 1.38\pm0.08 \nonumber \\
&&(\sigma = 0.63~~{\rm for}~L_{\rm bol}/L_{\rm Edd} > 0.01),
\label{pj2}
\end{eqnarray}
respectively. These two relations have similar slopes, with different intercepts. The overall linear fit gives that
\begin{eqnarray}
&&\log P_{\rm j}/L_{\rm Edd} = 0.48\pm0.02~\log L_{\rm bol}/L_{\rm Edd} - 1.44\pm0.04  \nonumber \\
&&(\sigma = 0.56).
\label{pj1}
\end{eqnarray}

The jet production efficiency is found to be negatively correlated with the Eddington ratio for radio galaxies~\citep{2015MNRAS.449..431J, 2013ApJ...764L..24S}. Here we add blazars, RL-NLS1s, and YRSs into this diagram (bottom panel of Figure~\ref{eddr_pj}). Blazars (mostly FSRQs) locate at the similar position with 3CR HEG/FR IIs, with higher Eddington ratio and lower jet production efficiency. FR Is in 3CR and FRICAT, and LEG/FR IIs in 3CR and FRIICAT show lower Eddington ratio and higher jet production efficiency. RL-NLS1s and YRSs do not follow the negative correlation between $L_{\rm bol}/L_{\rm Edd}$ and $P_{\rm j}/L_{\rm bol}$. YRSs with higher $L_{\rm bol}/L_{\rm Edd}$ have relatively higher $P_{\rm j}/L_{\rm bol}$, while those with lower $L_{\rm bol}/L_{\rm Edd}$ show lower $P_{\rm j}/L_{\rm bol}$. The jet production efficiency of RL-NLS1s is larger than FSRQs, while the Eddington ratio shows similar trend between FSRQs and RL-NLS1s. We also apply a linear regression between $L_{\rm bol}/L_{\rm Edd}$ and $P_{\rm j}/L_{\rm bol}$. The result gives that
\begin{eqnarray}
&&\log P_{\rm j}/L_{\rm bol} = -0.57\pm0.02~\log L_{\rm bol}/L_{\rm Edd} - 1.51\pm0.04 \nonumber \\
&&(\sigma = 0.55).
\label{eff1}
\end{eqnarray}
The relations for the sources with high and low Eddington ratios are
\begin{eqnarray}
&&\log P_{\rm j}/L_{\rm bol} = -0.64\pm0.08~\log L_{\rm bol}/L_{\rm Edd} - 1.80\pm0.25 \nonumber \\
&&(\sigma = 0.32~~{\rm for}~~L_{\rm bol}/L_{\rm Edd} < 0.01),\\
&&{\rm and} \nonumber \\
&&\log P_{\rm j}/L_{\rm bol} = -1.05\pm0.10~\log L_{\rm bol}/L_{\rm Edd} - 1.83\pm0.08 \nonumber \\
&&(\sigma = 0.57~~{\rm for} ~~L_{\rm bol}/L_{\rm Edd} > 0.01),
\label{eff2}
\end{eqnarray}
respectively. The slopes of these two population are slightly different. The sources with lower jet production efficiency show a steeper slope with Eddington ratio.

We note that all the linear relations above have large scatter ($>$ 0.3 dex) when all the subclasses of jetted AGNs, especially YRSs and RL-NLS1s, are included. This can be due to additional physical factors affecting jet power, such as evolution or the black hole spin (Section 5.3).

\section{Discussion}
\subsection{The Nature of LBLs}
FSRQs and BL Lacs are classified based on the strength of emission lines at optical band. FSRQs show prominent emission lines, while BL Lacs show very weak, even none line emission. Currently, most researches support that the observational difference is fundamentally due to the distinction on the accretion mode. FSRQs are powered by the radiatively efficient accretion disk associated with ionized emission line region. BL Lacs are fueled by hot gas and radatively inefficient accretion~\citep{2009MNRAS.396L.105G, 2011MNRAS.414.2674G, 2012MNRAS.420.2899G, 2014MNRAS.445...81S}. In another side, the synchrotron peak frequency of BL Lacs can span from $10^{12}$ Hz to as high as $10^{18}$ Hz. The properties of HBLs are usually found to be different with LBLs~\citep{1993ApJ...407...65G, 1995ApJ...444..567P}. LBLs are usually selected from radio surveys, thus they are also named radio-selected BL Lacs (RBLs) historically, while HBLs are mostly X-ray selected~\citep{1995PASP..107..803U, 2012MNRAS.420.2899G}. LBLs can show relatively stronger emission line and higher luminosity than typical HBLs~\citep{2010MNRAS.402..497G}.~\citet{2012MNRAS.420.2899G} suggested that LBLs and FSRQs were the same objects with different dilutions from jet emission at optical band. It was also found that there were some LBLs at high redshift with the radio morphologies similar with FR IIs~\citep{1995PASP..107..803U}.

The jet powers of LBLs span a wide distribution (Figure~\ref{3lac}). The K-S tests manifest that the high power LBLs ($>10^{44.6}$ erg s$^{-1}$) show similar jet power distribution with FSRQs, while the jet power distribution of low power LBLs is similar with that of IBLs (Table~\ref{cor}). These results indicate that not all, but fractional LBLs are intrinsically FSRQs. The other LBLs are indeed BL Lacs which host low power jet and featureless spectra. But it still needs to answer why these high power LBLs show weak emission lines than typical FSRQs. One possibility is that they have stronger Doppler boosting than FSRQs, which dilutes the emission lines~\citep{2012MNRAS.420.2899G}. But there is also another possibility that their central engines are intrinsically different, i.e., the aligned counterparts of LEG/FR IIs (see details in Section~\ref{unification}).

\subsection{The Nature of LEG/FR IIs}
The Eddington ratios of radio galaxies are clear bimodal at about 0.01, which correspond to HEGs and LEGs for high and low Eddington ratios, respectively~\citep{2010A&A...509A...6B, 2012MNRAS.421.1569B}. Thus HEGs and LEGs are suggested to be fundamental different on the accretion modes and the environments of host galaxies~\citep{2012MNRAS.421.1569B, 2016A&ARv..24...10T}. HEGs are fueled with radiatively efficient accretion, and hosted by galaxies with more cold gas and higher star formation rate. LEGs are associated with radiatively inefficient accretion. The host galaxies of LEGs are gas poor. FR Is are usually found to be LEGs. Among FR IIs, both HEGs and LEGs are obseved~\citep{2010A&A...509A...6B}. LEG/FR IIs are believed to be intrinsically different with the strong lined FR IIs in 3CR.~\citet{2013A&A...560A..81B} also showed that the difference between LEG/FR IIs and HEG/FR IIs in 3CR was not due to the orientation, but the accretion mode. Therefore, LEG and weak line FR IIs are taken as the outliers of unification model based on the orientations~\citep{1995PASP..107..803U, 2016A&ARv..24...10T}.

Our results manifest that the distributions of jet power and Eddington-scaled jet power are similar between LEG/FR IIs and HEG/FR IIs (Figure~\ref{rg} and Figure~\ref{pj_ledd_dis}), but the jet production efficiency of LEG/FR IIs is larger than HEG/FR IIs (Figure~\ref{pj_lbol_dis}). The jet production efficiencies of LEGs (including FR Is and LEG/FR IIs in 3CR) are larger than 1 (i.e., $P_{\rm j} > L_{\rm bol}$), while HEGs are on the contrary (Figure~\ref{pj_lbol_dis}). On the $L_{\rm bol}$ --- $M_{\rm BH}$ diagram, HEG/FR IIs and LEG/FR IIs are generally divided by the line $L_{\rm bol} = 0.01 L_{\rm Edd}$. These results suggest that the jet properties are similar for HEG/FR IIs and LEG/FR IIs. But their accretion properties are different, which leads to the difference on the jet production efficiency~\citep{2013ApJ...764L..24S, 2013ApJ...765...62S}.

\citet{2016A&ARv..24...10T} suggested that LEG/FR IIs originated from the intermittent activity or the evolutional consequence of the central engine, where the shut-down timescale of jet is much longer than the accretion disk and emission line region. About 10$^6$ year after switch-off-phase of accretion process, the jet activity would become weak, and the objects will be appeared as weak lined FR Is. This suggests an evolutional sequence among various classes of radio galaxies. HEG/FR IIs are triggered by merger events, where the central engines are fueled by the cold gas and radiatively efficient accretion. When the central engines switch off, the jets are still active, which are appeared as LEG/FR IIs. After the jet activities also get weak, they will act as FR Is, which can be fueled by the hot gas and radiatively inefficient accretion~\citep{2010A&A...509A...6B, 2012MNRAS.421.1569B, 2016A&ARv..24...10T}. This scenario expects that the gas environments are similar between LEG/FR IIs and FR Is, as their accretion process might be both supplied by the hot gas. Actually~\citet{2017A&A...601A..81C} noted that the LEG/FR IIs in FRIICAT shared similar properties of black hole mass and host galaxy with FR Is in FRICAT.

\subsection{The Unification of Jetted AGNs}
\subsubsection{The Unification between Blazars and Radio Galaxies}
\label{unification}
Blazars and radio galaxies (at least for FR Is and strong-lined FR IIs) are believed to be the AGNs with same properties. The observational differences are due to the anisotropic radiation caused by Doppler beaming effect~\citep{1995PASP..107..803U}. This unification scenario between blazars and radio galaxies has been verified in the literatures by many authors~\citep{2001ApJ...548..244B, 2009ApJ...694L.107X, 2015MNRAS.451.2750X}. In this paper, we compare the jet powers (extended jet emission) for BZCAT blazars and 3CR radio galaxies. The results of K-S test for the jet power distributions support the unifications between FR Is/FR IIs and BL Lacs/FSRQs, respectively (Table~\ref{cor}). In opposite, if the unification scenarios is correct~\citep{1995PASP..107..803U}, the correction factor between the observed luminosity and the extended luminosity at 150 MHz for blazars is reliable~\citep{2018ApJ...869..133F}. Note that the significance of K-S test can be weakened by several selection effects~\citep{1995PASP..107..803U}, e.g., the different ranges of redshift for radio galaxies and blazars. Another scatter comes from the constant correction factor applied for whole blazar population (2.1 in our case,~\citealt{2018ApJ...869..133F}).

The distributions of jet production efficiency for HEG/FR IIs and FSRQ is slightly different, with the $P_{\rm j}/L_{\rm bol}$ of HEG/FR IIs larger than FSRQs (Table~\ref{cor}). The jet production efficiency is suggested to be related to the accretion mode, as the later affects the magnetic field~\citep{1997MNRAS.292..887G}. The radiative efficient accretion corresponds to low $P_{\rm j}/L_{\rm bol}$, while the high $P_{\rm j}/L_{\rm bol}$ is only observed in the source with low Eddington ratio (except YRSs, bottom panel of Figure~\ref{eddr_pj}). Thus the difference on the jet production efficiency between FSRQs and HEG/FR IIs could be caused by the potential evolution of accretion process from high and low redshifts (e.g.~\citealt{2016A&ARv..24...10T}). The radio galaxies in this work is confined by a redshift limit of z $<$ 0.3~\citep{2010A&A...509A...6B, 2011A&A...525A..28B, 2017A&A...598A..49C, 2017A&A...601A..81C}. This range of redshift is general consistent with BL Lacs, but much smaller than FSRQs. The evolution of blazars/radio galaxies from high to low redshifts needs more explorations.

Our results and several other works suggest that the high power LBLs are actually FSRQs, which are unified with FR IIs. But LBLs usually show weak emission lines than typical FSRQs. Among FR IIs, there are also some transitional objects with low excitation or weak line properties, i.e., LEG/FR IIs. This gives a hint that these high power LBLs may be unified with LEG/FR IIs. This scenario can be examined by the spectroscopy observation at infrared, as suggested by~\citet{2012MNRAS.420.2899G}. The number counts for flux-limited samples are also helpful. But it needs to keep in mind that the BL Lacs lacking redshift measurements would affect the results of population statics~\citep{2012MNRAS.420.2899G}.

In addition, the radio galaxies in FRICAT and FRIICAT show much lower jet power than blazars and 3CR radio galaxies. The low power radio galaxies show much smaller black hole mass compared with the 3CR radio galaxies (Table~\ref{cor}). They could be the low power counterparts with lower black hole mass compared with the classical radio galaxies. However, the Eddington-scaled jet power of FR IIs in FRIICAT (mostly LEGs) is slightly smaller than the LEG/FR IIs in 3CR, while their distributions of jet production efficiency are similar. On the other side, the FR Is in FRICAT and 3CR show similar distributions on the Eddington-scaled jet power. But the distributions of their jet production efficiency are different. These results indicate that some additional factors affect the jet power properties between low power and classical radio galaxies.

Until now, we do not consider the effect on the jet power due to different black hole spins~\citep{2007ApJ...658..815S}.~\citet{2008ApJ...687..156W} calculated the jet power for radiative inefficient accretion flow with different black hole spins based on a hybrid model~\citep{1999ApJ...522..753M}. The jet power can differ by one magnitude of order for different black hole spins with same accretion rate (also see~\citealt{2013ApJ...764L..24S}). Thus, the different black hole spins can be a possible reason to explain the low power of radio galaxies in FRICAT and FRIICAT.

According to the unification model, there should be a population of aligned low power radio sources, which is unified with the radio galaxies in FRICAT and FRIICAT. They can be low power BL Lacs and FSRQs, which host lighter black hole than typical blazars and radio galaxies. Another possible candidate is the flat-spectrum RL-NLS1~\citep{2008ApJ...685..801Y, 2015A&A...575A..13F, 2015A&A...578A..28B, 2016A&A...591A..98B}. However, the jet properties (including jet power, Eddington-scaled jet power, and jet production efficiency) and central engine properties (including bolometric luminosity, black hole mass and Eddington ratio) are all different between RL-NLS1s and FR IIs in FRIICAT (Table~\ref{cor}). In addition, the radio morphologies of RL-NLS1s are often found to be compact~\citep{2010AJ....139.2612G, 2015ApJS..221....3G}. RL-NLS1s also show strong emission lines, which is different with the LEG properties of FRICAT and FRIICAT.

We stress that it still lacks detailed information on black hole mass and accretion property for a substantial sample of BL Lacs, which prevent a detailed comparison for the jet-disk connection between FR Is and low power blazars, as well as between LBLs and LEG/FR IIs. BL Lacs with emission lines are found to own similar properties with FSRQs~\citep{2014MNRAS.445...81S}. For lineless BL Lacs, the bolometric luminosity, even the black hole mass, is difficult to be constrained. Some attempts are based on the SED fit with a standard disk model~\citep{2010MNRAS.402..497G}. As BL Lacs are believed to be powered by radiatively inefficient accretion, we do not include these samples in our work.

\subsubsection{The Unification between RL-NLS1s and Young Radio Sources}
The discover of $\gamma$-ray NLS1s presents a problem that what is the parent population of them. Steep-spectrum RL-NLS1s are nature candidates. But the number of steep-spectrum RL-NLS1s are much less than the expected from the aligned estimation~\citep{2016A&A...591A..98B}. Thus some authors suggested YRSs as the parent population of $\gamma$-ray NLS1s (or flat-spectrum RL-NLS1s,~\citealt{2015A&A...575A..13F, 2016A&A...591A..98B}).~\citet{2016A&A...591A..98B} investigated the luminosity function and properties of central engines, and concluded that HEG/CSSs could be the aligned counterparts of flat-spectrum RL-NLS1s. Limited by their sample size, more explorations are needed.

The jet powers of YRSs in our samples show a very broad distribution from 10$^{42}$ erg s$^{-1}$ to 10$^{47}$ erg s$^{-1}$, while the jet powers of RL-NLS1 are generally less than 10$^{45}$ erg s$^{-1}$. The Eddington-scaled jet power of RL-NLS1s is larger than that of YRSs, while the jet production efficiency is on the opposite. The black hole masses of young radio sources are larger than RL-NLS1s, which is similar with the suggestions of~\citet{2017NewA...50...78F} based on a much smaller sample. The mean Eddington ratio of RL-NLS1s is larger than that of YRSs (Table~\ref{cor}). These results suggest that YRSs are not the unified version of flat-spectrum RL-NLS1s, although the subclasses of YRSs, i.e., low power HEG/CSSs can be a possible candidate~\citep{2016A&A...591A..98B,2018rnls.confE..26B}. Combined with the difference on the propertied of host galaxies between YRSs and RL-NLS1s~\citep{1998PASP..110..493O, 2015A&A...575A..13F}, the unification between these two types of jetted AGNs are more complicated than the simple orientation effect. There seems the black hole growth and galaxy evolution also take an role.

The jet production efficiency of YRSs is generally equal to 1, similar with the sources of low Eddington ratio. On the $P_{\rm j}/L_{\rm bol}$ --- $L_{\rm bol}/L_{\rm Edd}$ plane, YRSs are obviously incompatible with the negative correlation between these two parameters (bottom panel of Figure~\ref{eddr_pj}). The eccentric high jet production efficiency of YRSs has been noted in the previous work (\citealt{2009MNRAS.398.1905W}; Liao et al. 2019). This feature can be seen as an evidence of the early-stage jet activity for YRSs. RL-NLS1s are also accepted to be the evolved version of AGNs. Their jet production efficiency is also larger than that of FSRQs, but lower than that of YRSs (Table~\ref{cor}).

\subsection{Black Hole Mass and the Evolution of Jetted AGNs}
\begin{figure}
\includegraphics[angle=0,scale=.4]{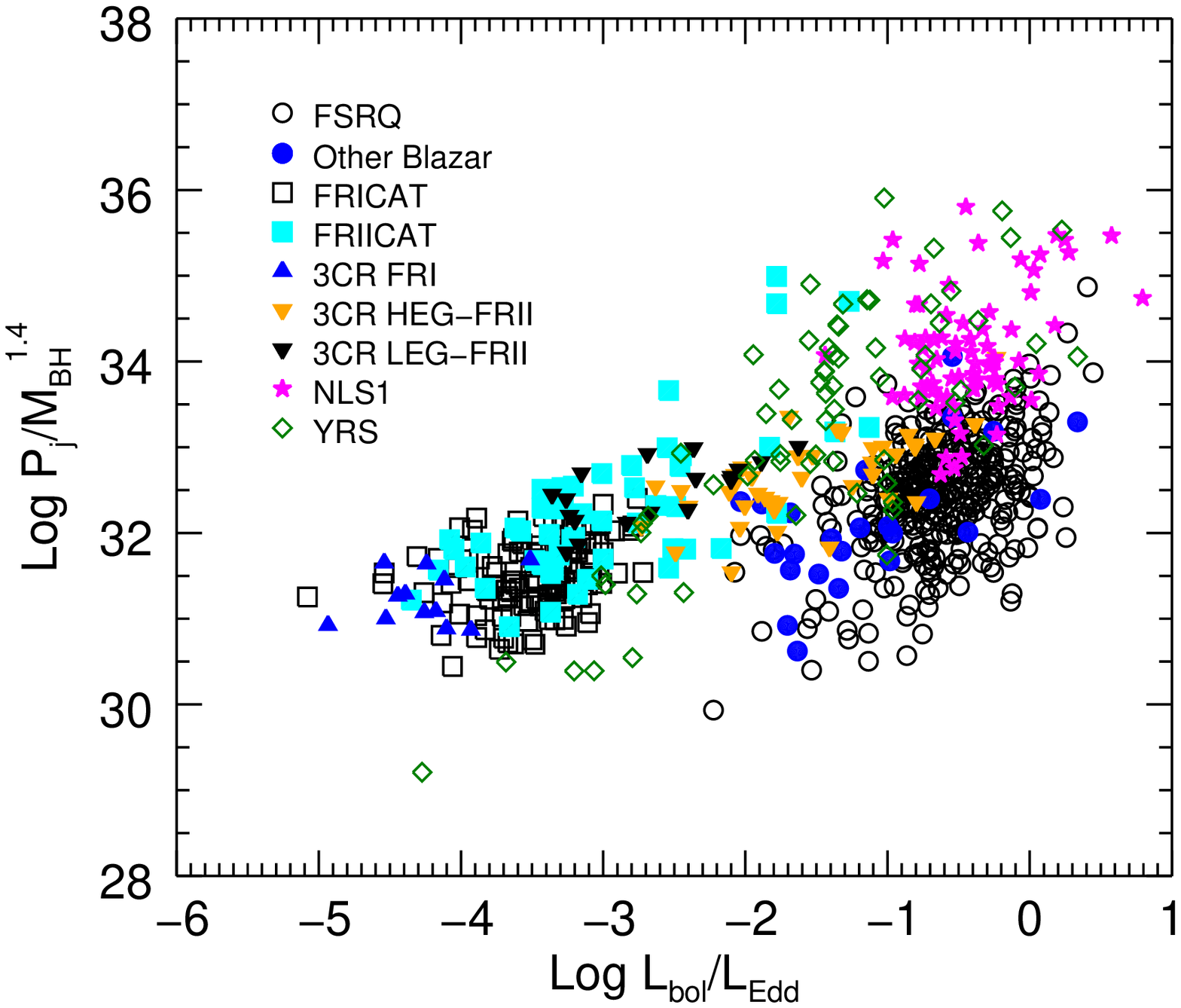}
\includegraphics[angle=0,scale=.4]{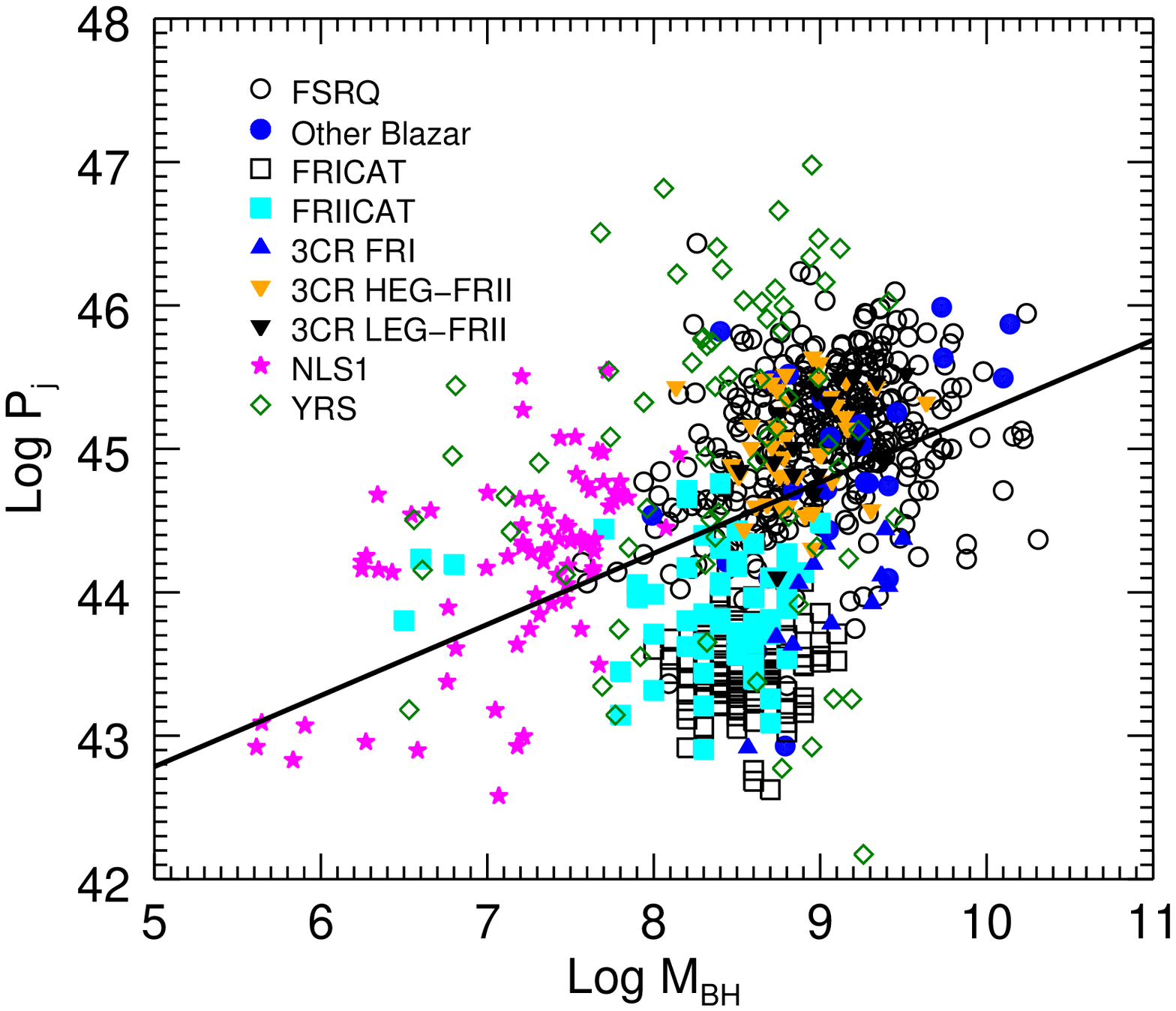}
\caption{Top panel: The Eddington ratio versus normalized jet power of jetted AGNs. Bottom panel: The jet power against the black hole mass for jetted AGNs. The black solid line shows the best fit. The representation of each symbol is labelled in the figures.  \label{mass_pj}}
\end{figure}

It is expected that high mass black hole will result in high power jet, at least for the sources with same accretion mode. Several authors suggested that black hole systems had similar jet power if the black hole mass was normalized~\citep{2017arXiv170305575F, 2019ApJ...872..169P}. We have divided the jet power with Eddington luminosity, and find that different types of jetted AGNs show different ranges of $P_{\rm j}/L_{\rm Edd}$ (Figure~\ref{pj_ledd_dis} and Figure~\ref{eddr_pj}). In addition, there is also suggestion that jet power scales with black hole mass with $P_{\rm j} \propto M_{\rm BH}^{1.4}$, which is similar for both radiatively efficient and inefficient accretions (for radiatively inefficient accretion, jet power also scales with accretion rate, i.e., $P_{\rm j} \propto (M_{\rm BH}\dot{m})^{1.4}$,~\citealt{2003MNRAS.343L..59H}). Following the suggestion of~\citet{2017arXiv170305575F}, we explore the distribution of $P_{\rm j}/M_{\rm BH}^{1.4}$ (hereafter normalized jet power, top panel of Figure~\ref{mass_pj}) for our samples. Different types of sources also show different ranges of normalized jet power. RL-NLS1s and fractional YRSs still have the largest normalized jet powers. FR Is have the smallest normalized jet powers. The normalized jet power of low Eddington ratio objects show very different behaviors with the higher Eddington ratio ones. For the high Eddington ratio sources, the normalized jet power are increased with the Eddington ratio. For low Eddington ratio sources, this trend is very weak.

We also fit the relation between jet power and black hole mass. The result is plotted in the bottom panel of Figure~\ref{mass_pj}. The linear relation gives that $P_{\rm j} \propto M_{\rm BH}^{0.50\pm0.05}$ with a large scatter of 0.73 dex.

RL-NLS1s are believed as the low black hole mass version of FSRQs and HEG/FR IIs~\citep{2009ApJ...707L.142A}. We compare the properties between these two types of sources. The Eddington-scale jet power, normalized jet power and jet production efficiency are all different between RL-NLS1s and FSRQs (Table~\ref{cor}), which indicates that the jet activities of these two types of sources are also different besides the black hole mass. The evolution of accretion and environment seems important in addition to the black hole growth.

\section{Summary and Conclusion}
Jet power is a fundamental parameter of jet physics. It is important to distinguish and unify different subclasses of jetted AGNs. In this paper, we estimate the jet power for various samples of jetted AGNs with the low-frequency radio data at 150 MHz from TGSS AD1. With these jet power estimations, we firstly compare the jet power among different subclasses of blazars. The jet power of LBLs shows a bimodal distribution with a transition point at about $10^{44.6}$ erg s$^{-1}$. This suggests there are two types of LBLs. The first is indeed BL Lacs with low jet power and featureless optical spectra. The other is FSRQs which are classified as BL Lacs due to some intrinsic differences, or selection effects, e.g., the dilutions due to the beaming emission from jet (e.g.~\citealt{2012MNRAS.420.2899G}).

After correcting the contribution from radio core of blazars, the unification between blazars and radio galaxies is confirmed with the jet power distributions. There are also a new class of low power radio galaxies with lower black hole mass than typical blazars and radio galaxies.

The optical and accretion properties of nearby radio galaxies have been wildly explored (see the review of~\citealt{2016A&ARv..24...10T}). For other jetted AGNs, the accretion properties are hard to constrain due to the overwhelming jet contribution (blazars, especially lineless BL Lacs), or limited sample sizes (RL-NLS1s and YRSs). In this paper, we explore the jet-disk connection combined various samples of jetted AGNs. Our results manifest that the jetted AGNs can be divided into two populations on the Eddington ratio --- Eddington-scaled jet power plane. This division seems related to the accretion modes. As the jet power of high Eddington ratio sources is less than the accretion power, while the low Eddington ratio sources show the opposite trend. The jet production efficiency of jetted AGNs is also important to distinguish their subclasses and intrinsic properties, such as the accretion modes. Jetted AGNs follow similar negative correlation between Eddington ratio and jet production efficiency.

HEG and LEG FR IIs are clearly separated by the equation line on the Eddington ratio --- Eddington-scaled jet power diagram, and by the line $L_{\rm bol} = 0.01~L_{\rm Edd}$ on the $P_{\rm j}$ --- $M_{\rm BH}$ plane. We suggest an evolutional sequence of accretion process and jet activity among various classes of radio galaxies, which evolves from HEG/FR IIs, LEG/FR IIs to FR Is. The high power LBLs ($> 10^{44.6}$ erg s$^{-1}$) might unify with LEG/FR IIs.

The Eddington-scaled jet power and jet production efficiency of NLS1s and FSRQs are not consistent with each other. These results suggest that the unification between NLS1s and other jetted AGNs is more complicated rather than the black hole growth from smaller black hole to heavier ones. In addition, the unification between NLS1s and YRSs is not confirmed based on the properties of jet and central engines.

YRSs are the obviously outliers on the Eddington ratio --- Eddington-scaled jet power plane and Eddington ratio --- jet production efficiency palne. They show higher jet power than accretion power with relatively high Eddington ratio. They also show higher jet production efficiency than other objects with high Eddington ratio.

\acknowledgments
We are grateful to the anonymous referee for the suggestive comments. We thank Mai Liao for kindly sending us the data of young radio sources. This research is supported by National Natural Science Foundation of China (NSFC; grants 11573009 and 11622324) and by Project funded by China Postdoctoral Science Foundation (2018M642811).

%

\vspace{5mm}
\facilities{GMRT (TGSS AD1), Fermi/LAT (3LAC)}


\bibliographystyle{aasjournal}
\bibliography{bib}

\end{document}